%% file: ae2_new.tex
\documentstyle[11pt,epsfig,url,graphicx,hyperref]{article}
\hypersetup{colorlinks=true, linkcolor=blue, filecolor=blue,
pagecolor=blue, urlcolor=blue}
        \catcode`\@=11
        \@addtoreset{equation}{section}

\def\MZ{M_Z}

\def\dd{.\kern-1pt.\kern-1pt.}
\def\dahz{\Delta\alpha^{(5)}_{\rm had}(\MZ^2)}
\def\dah0{\Delta\alpha^{(5)}_{\rm had}(-M_0^2)}

\newcommand{\Dafne}{DA$\Phi$NE}

\def\ifm#1{\relax\ifmmode#1\else$#1$\fi}
 
\def\to{\ifm{\rightarrow}} 
  
\def\up#1{$^{#1}$}    
  
\def\ie{{\it\kern-2pt i.\kern-.5pt e.\kern-2pt}}  
\def\bye{\end{document}}
\def\dif{\hbox{d\kern.1mm}}

\def\up#1{$^{#1}$}  

\makeatletter
\def\@listi{\leftmargin\leftmargini
            \parsep .1\parskip
            \topsep \parsep
            \itemsep .1\parskip
            \partopsep \z@}
\makeatother

\begin{document}

\vspace{1cm}
\begin{flushright}
{\bf LNF-10/17(P)} \\
\end{flushright}
\vspace{1cm}

\begin{center}
\vspace{1.5cm}
{\LARGE \bf
Proposal for taking data with the KLOE-2\\[1mm]
 detector at the \Dafne\ collider 
upgraded in energy}

\vspace{1.0cm}
%
\vspace{0.5cm}
D.~Babusci{\up a}, C.~Bini{\up b}, F.~Bossi{\up a}, G.~Isidori{\up a}, D.~Moricciani{\up c}, F.~Nguyen{\up d}, P.~Raimondi{\up a}, G.~Venanzoni{\up a}, 
D.~Alesini{\up a}, F.~Archilli{\up c}
% ...
 

\vspace{1.5cm}
{\up a}{\it Laboratori Nazionali di Frascati dell'INFN, Frascati, Italy} \\
{\up b}{\it Dipartimento di Fisica dell'Universit\`a ``La Sapienza''
e Sezione INFN, Rome, Italy} \\
{\up c}{\it Dipartimento di Fisica dell'Universit\`a ``Tor Vergata''
e Sezione INFN, Rome, Italy}\\
{\up d}{\it Dipartimento di Fisica dell'Universit\`a ``Roma Tre''
e Sezione INFN, Rome, Italy} \\
{\up e}{\it
Institute for Theoretical and Experimental Physics, Moscow, Russia} \\
{\up f}{\it  Dipartimento di Fisica dell'Universit\`a di Bari 
e Sezione INFN, Bari, Italy} \\
{\up g}{\it INFN, Sezione di Napoli, Napoli, Italy} \\
{\up h}{\it University of Uppsala, Uppsala, Sweden} \\
{\up i}{\it  Dipartimento di Energetica, Universit\`a Sapienza, Rome, Italy} \\
{\up j}{\it Institut f\"ur Experimentelle Kernphysik, Universit\"at Karlsruhe, Karlsruhe, Germany} \\
{\up k}{\it Budker Institute of Nuclear Physics, Novosibirsk, Russia} \\
{\up l}{\it Universit\`a della Calabria, Cosenza, e  INFN Gruppo collegato di Cosenza, Cosenza, Italy} \\
{\up m}{\it Jagiellonian University, Cracow, Poland} \\
{\up n}{\it Institut f\"ur Kernphysik, Johannes Gutenberg-Universit\"at Mainz, Mainz, Germany} \\
{\up o}{\it INFN, Sezione di Padova, Padova, Italy} \\
{\up p}{\it CERN, Geneve, Switzerland} \\
{\up q}{\it A. Soltan Institute for Nuclear Studies, Warsaw, Poland}\\
{\up r}{\it Museo Storico della Fisica e Centro Studi e Ricerche 
``E. Fermi'', Rome, Italy}




\end{center}

\begin{abstract}
This document reviews the physics program of the KLOE-2 detector at
\Dafne\ upgraded in energy and provides a simple solution to run 
the collider above the $\phi$-peak (up to 2, possibly 2.5 GeV).
 It is shown how a precise measurement of the multihadronic 
cross section in the energy region up to 2 (possibly 2.5) GeV would have  
a major impact on the tests of the Standard Model through a precise 
determination of the anomalous magnetic moment of the muon and the 
effective fine-structure constant at the $M_Z$ scale.
With a luminosity of about $10^{32}$cm$^{-2}$s$^{-1}$, 
DA$\mathrm{\Phi}$NE upgraded in energy 
can perform a scan in the region from 1 to 2.5 GeV in one year by collecting
an integrated luminosity  of 20 pb$^{-1}$ (corresponding
to a few days of data taking) for single point,
 assuming an energy step of 25 MeV.
A  few years of data taking in this region would provide 
important tests of  QCD and effective theories
by  $\gamma\gamma$ physics with open 
thresholds for pseudo-scalar 
(like the $\eta'$), scalar ($f_0,f'_0$, etc...) and 
axial-vector ($a_1$, etc...) mesons;
vector-mesons spectroscopy and baryon form factors; tests of CVC
 and searches for exotics. 
In the final part of the document a technical
solution for the energy upgrade of \Dafne\ is proposed.
\end{abstract}
\clearpage

\tableofcontents

\section{Introduction}
\label{intro}
\input{intro_fb}

\section{Precision tests of the Standard Model}
\label{subsubsec:SMTESTS}
\input{testSM}

\section{Measurement of the hadronic cross sections below 2.5 GeV}
\label{hadc}
\input{hadc}


\section{Other physics motivations}
\label{phys}
\input{phys}


\section{\Dafne\ Energy Upgrade}
\label{dafne}
\input{dafne}

\section{Summary}
Precision tests of the Standard Model
in future experiments require a more accurate knowledge of the hadronic
cross section in the whole energy range between the  2$m_{\pi}$ threshold
 and 2.5 GeV. The region between 1 and 2.5 GeV is at
present the most poorly known and is crucial for the computation
of the hadronic corrections to the effective 
fine-structure constant at the $M_Z$ scale, $\alpha_{em}(M^2_Z)$.
It rapresents also a limiting factor to the accuracy of the SM prediction of the 
$g-2$ of the muon.
With an energy upgrade of the \Dafne\ collider,
KLOE-2 can reduce the accuracy of the hadronic contribution of the muon anomaly
$a_{\mu}^{\rm HLO}$
to less than $3\times$10$^{-10}$.
This would represent a twofold reduction of the present error, 
 necessary to match the increased precision of the proposed muon $(g-2)$ experiments at FNAL and J-PARC, and to firmly establish (or constrain) 
``new physics'' effects. 
A similar improvement can be expected on the determination of the hadronic contribution to $\alpha_{em}(M^2_Z)$.
Additional motivations for this upgrade are provided by  tests of  QCD and effective theories by  $\gamma\gamma$ physics with open 
thresholds for the production of pseudo-scalar
(like the $\eta'$), scalar ($f_0, a_0, f'_0, a'_0$, etc...) and 
axial-vector ($a_1, f_1, a'_1, f'_1$, etc...) mesons;
vector-meson spectroscopy and baryon form factors; tests of CVC
 and searches for exotics.
A technical solution for the energy upgrade of \Dafne\ (up to about 2 GeV) 
is proposed.

\section*{Acknowledgements}
We would like to thank Henryk Czy\.z, Simon Eidelman, Fred Jegerlehner and Andreas Nyffeler
 for useful discussions.
\bibliographystyle{epj}
\bibliography{ae}


\end{document}

%% file: intro_fb.tex
In this document we discuss the physics program that can be pursued 
by running DA$\mathrm{\Phi}$NE at energies above the $\phi$ meson peak,
 and the upgrades to the 
machine required for this purpose 
(the so called DA$\mathrm{\Phi}$NE-2 program).

We consider  a reference luminosity of 10$^{32}$cm$^{-2}$s$^{-1}$, 
from the $\phi$ up to $\sim$2.5 GeV, which seems to be feasible with a 
modification of the existing machine 
at a moderate cost,
 as explained in the following. 
With such a machine one can easily collect an integrated luminosity of 
 about 5 fb$^{-1}$ between 1 and 2.5 GeV in a few years of data taking.
This high statistics,  much larger than that collected at
 any previous machine in this energy range, 
will represent a major improvement in physics, with 
relevant implications for the precision tests of the Standard Model, like the  
$g-2$ of the muon and the effective fine-structure constant at the $M_Z$ scale 
 $\alpha_{em}(M^2_Z)$. The only direct competitor project is VEPP-2000 at Novosibirsk, 
which will cover
the center-of-mass energy range between 1 and 2 GeV with two experiments. This
collider has started first operations in 2009 and is expected  to provide a luminosity ranging between $10^{31}$cm$^{-2}$s$^{-1}$ at 1 GeV and $10^{32}$cm$^{-2}$s$^{-1}$ at 2 GeV.
Other ``indirect'' competitors are the higher
energy $e^+e^-$ colliders ($\tau$-charm and B-factories) that in principle
cover the DA$\mathrm{\Phi}$NE-2 energy range by means of radiative return.
However, due to the photon emission, as we will show later, 
the ``equivalent'' luminosity produced by these machines
 in the region between 1 and 2.5 GeV is
much less than the one expected by DA$\mathrm{\Phi}$NE-2. 

The KLOE detector has succesfully taken data on  DA$\mathrm{\Phi}$NE in the last ten years~\cite{Bossi:2008aa}.
It is presently starting a new period of data taking at the $\phi$ peak, with some 
hardware modification either already implemented or planned to be implemented 
in  2011 (so called KLOE-2 project)~\cite{kloe2paper,AmelinoCamelia:2010me}. 
Although a detailed Monte Carlo simulation has not been carried out yet,
its measured performance, together with the improvements expected from the 
insertion of the new subdetectors, make us confident that 
KLOE-2 is the proper detector for this kind of measurements.
Actually, data taking at energies higher than the $\phi$ mass is a relevant
part of the KLOE-2 physics program.
  
In the following sections we first present the main physics motivations for this high-energy program~\cite{kloe2paper,AmelinoCamelia:2010me,roadm}.
 We start with the implication for precision tests of the Standard Model
from a precise measurement 
 of the multi-hadronic cross section in the energy region below 2.5 GeV.
In particular, we discuss the strategies for a precise determination
of the effective fine structure constant at the scale $M_{Z}$, and of the 
muon $g-2$. We then concentrate on the potential of the proposed machine
for the measurement of this cross section, comparing it with possible 
competitors. Other physics topics that
can benefit from  DA$\mathrm{\Phi}$NE-2 are briefly discussed in Section~\ref{phys}. 
Finally, we discuss the main technical issues to be addressed to properly
modify the machine for our purpose. 
%

%% file: testSM.tex
The systematic comparison of Standard Model ({\small SM})
predictions with precise experimental data served in the last 
decades as an invaluable tool to test this theory at the quantum
level. It has also provided stringent constraints on ``new
physics'' scenarios.  The (so far) remarkable agreement between the
measurements of the electroweak observables and their {\small
  SM} predictions is a striking experimental confirmation of the
theory, even if there are a few observables where the agreement is not
so satisfactory.  On the other hand, the Higgs boson has not yet been
observed, and there are clear phenomenological facts (dark matter, 
matter-antimatter asymmetry in the universe) as well as 
strong theoretical arguments hinting at the
presence of physics beyond the {\small SM}. The LHC, 
 or a future $e^+ e^-$ International Linear Collider
(ILC), will hopefully answer many questions.
 However, 
their discovery potential may be substantially improved if 
combined with more precise low-energy tests of the SM.
\subsection{The effective fine-structure constant at the scale $M_Z$}
\label{subsubsec:ALPHAEFF}
\input{alpha_gv}

\subsection{The muon $g$$-$$2$}
\label{subsubsec:GMINUS2}
\input{g-2_cb2}

%% file: alpha_gv.tex
Precision tests of the Standard Model require the appropriate inclusion
of higher order effects and the knowledge of very precise input parameters.
One of the basic input parameters is the fine-structure constant 
$\alpha$, determined from the  anomalous 
 magnetic moment of the electron
 with an impressive accuracy of 0.37 parts per billion 
(ppb)~\cite{Hanneke:2008tm} 
relying on the validity of 
 perturbative QED~\cite{Gabrielse:2006gg}.
However, physics at nonzero squared momentum transfer $q^2$
is actually described by an effective electromagnetic
coupling $\alpha(q^2)$ rather than by the low-energy
constant $\alpha$ itself. The shift of the fine-structure
constant from the Thomson limit to high energy involves low
energy non-perturbative hadronic effects which spoil this
precision. In particular, the effective fine-structure
constant at the scale $M_{\scriptscriptstyle{Z}}$,
$\alpha(M_{\scriptscriptstyle{Z}}^2) = \alpha/[1-\Delta
\alpha(M_{\scriptscriptstyle{Z}}^2)]$, plays a crucial role
in basic electroweak {\small EW} radiative corrections of the {\small
  SM}. An important example is the {\small EW} mixing
parameter $\sin^2 \!\theta$, related to $\alpha$, the Fermi
coupling constant $G_F$ and $M_{\scriptscriptstyle{Z}}$ via
the Sirlin relation~\cite{Sirlin:1980nh,Sirlin:1989uf,Marciano:1980pb}
\begin{equation}
  \sin^2 \!\theta_{\scriptscriptstyle{S}} \cos^2 \!\theta_{\scriptscriptstyle{S}} = 
  \frac{\pi \alpha}
{\sqrt 2 G_F M_{\scriptscriptstyle{Z}}^2 (1-\Delta r_{\scriptscriptstyle{S}})},
\label{eq:sirlin}
\end{equation}
where the subscript $S$ identifies the renormalisation
scheme. $\Delta r_{\scriptscriptstyle{S}}$ incorporates the
universal correction $\Delta
\alpha(M_{\scriptscriptstyle{Z}}^2)$, large contributions
that depend quadratically on the top quark
mass $m_t$~\cite{Veltman:1977kh}, plus all remaining
quantum effects.
\begin{figure}[ht]
\begin{center}
\includegraphics[width=6cm]{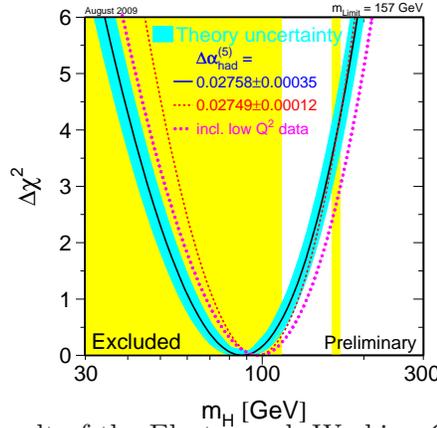}
\vspace{-1.cm}
\caption{The line is the result of the Electroweak Working Group fit
  using all data (see ~\cite{LEPEWWG} for details); the (blue) band represents an estimate of
  the theoretical error due to missing higher order corrections.  The
  (yellow) vertical bands show the 95\% CL exclusion limits on
  $M_{\scriptscriptstyle{H}}$ from the direct searches.}
\label{fig:blueband}
\end{center}
\end{figure}

In the {\small SM}, $\Delta r_{\scriptscriptstyle{S}}$
depends on various physical parameters, including
$M_{\scriptscriptstyle{H}}$, the mass of the Higgs boson. As
this is the only relevant unknown parameter in the {\small
  SM}, important indirect bounds on this missing ingredient
can be set by comparing the calculated quantity in
Eq.~(\ref{eq:sirlin}) with the experimental value of $\sin^2
\!\theta_{\scriptscriptstyle{S}}$ (e.g.\ the effective
{\small EW} mixing angle $\sin^2 \!\theta_{\rm eff}^{\rm
  lept}$ measured at LEP and SLC from the
on-resonance asymmetries) once
$\Delta\alpha(M_{\scriptscriptstyle{Z}}^2)$ and other
experimental inputs like $m_t$ are provided. It is important
to note that the uncertainty of 
the effective
electromagnetic coupling constant 
$\delta
\Delta\alpha(M_{\scriptscriptstyle{Z}}^2)$
affects the upper bound for $M_H$~\cite{leplsd:2005ema,LEPEWWG,Passera:2008jk}, see Fig.~\ref{fig:blueband}.  Moreover, as
measurements of the effective {\small EW} mixing angle at a
future linear collider may improve its precision by one
order of magnitude, a much smaller value of
$\delta\Delta\alpha(M_{\scriptscriptstyle{Z}}^2)$ will be
required (see below). It is therefore crucial to assess all
viable options to further reduce this uncertainty.

\begin{figure}[t]
\centering
\includegraphics[height=6.7cm]{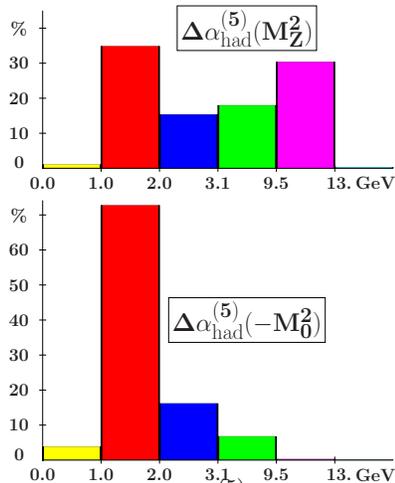}
\vspace*{-6mm}
\caption{Present error profiles for $\dahz$ ({\it standard integration, up}), and $\dah0$ ({\it Adler function, down}). As it can be seen with this second method about $70\%$ of the total error comes from the region below 2 GeV.}
\label{fig:errorprof} 
\end{figure}

The shift $\Delta \alpha(M_{\scriptscriptstyle{Z}}^2)$ can
be split in two parts:
$\Delta\alpha(M_{\scriptscriptstyle{Z}}^2) =
\Delta\alpha_{\rm lep}(M_{\scriptscriptstyle{Z}}^2) + \Delta
\alpha_{\rm had}^{(5)}(M_{\scriptscriptstyle{Z}}^2)$. The former one,
the leptonic contribution, is calculable in perturbation theory
and known up to three-loop accuracy: $\Delta\alpha_{\rm
  lep}(M_{\scriptscriptstyle{Z}}^2) = 3149.7686\times
10^{-5}$~\cite{Steinhauser:1998rq}. The hadronic contribution $\Delta
\alpha_{\rm had}^{(5)}(M_{\scriptscriptstyle{Z}}^2)$ of the
five light quarks ($u$, $d$, $s$, $c$, and $b$) can be
computed from hadronic $e^+ e^-$ annihilation data via the
dispersion relation~\cite{Cabibbo:1961sz}
\begin{equation} 
  \Delta \alpha_{\rm had}^{(5)}(M_{\scriptscriptstyle{Z}}^2) = 
-\left(\frac{\alpha M_{\scriptscriptstyle{Z}}^2}{3\pi}
  \right) \mbox{Re}\int_{m_\pi^2}^{\infty} {\rm d}s 
\frac{R(s)}{s(s- M_{\scriptscriptstyle{Z}}^2
  -i\epsilon)},
\label{eq:delta_alpha_had}
\end{equation}
where $R(s) = \sigma^{0}_{\rm had}(s)/(4\pi\alpha^2\!/3s)$ and
$\sigma^{0}_{\rm had}\!(s)$ is the total cross section for $e^+ e^-$
annihilation into any hadronic states, with 
vacuum polarisation and initial state {\small QED}
corrections subtracted off. The current
accuracy of this dispersion integral is of the order
of 1\%, dominated by the error of the hadronic cross section
measurements in the energy region below a few GeV~\cite{Eidelman:1995ny,Davier:1997kw,Burkhardt:2001xp,Burkhardt:2005se,Jegerlehner:2001wq,Jegerlehner:2003rx,Jegerlehner:2006ju,Jegerlehner:2008rs,Jegerlehner:2003qp,Jegerlehner:2003ip,Hagiwara:2003da,Hagiwara:2006jt} (see Fig.~\ref{fig:errorprof} {\it up}).
Table~\ref{tab:future} (from Ref.~\cite{Jegerlehner:2001wq}, updated in~\cite{Jegerlehner:2008rs}) shows that an
uncertainty $\delta \Delta\alpha_{\rm had}^{(5)} \sim 5 \times
10^{-5}$, needed for precision physics at a future linear collider,
requires the measurement of the hadronic cross section with a
precision of $1\%$ from threshold up to the $\Upsilon$ peak.
\begin{table}[h]
\begin{center}
 \renewcommand{\arraystretch}{1.4}
 \setlength{\tabcolsep}{1.6mm}
{\footnotesize
\begin{tabular}{|c|c|c|c|}
\hline
$\delta \Delta\alpha_{\rm had}^{(5)} \!\times \! 10^{5} $ 
& $\delta(\sin^2 \!\theta_{\rm eff}^{\rm lept}) \! \times \! 10^{5}$  
& Request on $R$\\
\hline \hline
22   &  7.9 & Present~\cite{Jegerlehner:2008rs} \\
\hline
7   &   2.5 & $\!\delta R/R \leq  1\%$ up to ${J/\psi}$\\
\hline   
5   &   1.8 & $\delta R/R \leq 1\%$ up to $\Upsilon$\\
\hline   
\end{tabular}
}
\caption{\label{tab:future} 
Values of the uncertainties $\delta
\Delta\alpha_{\rm had}^{(5)}$ (first column) and the errors induced by these
uncertainties on the theoretical {\small SM} prediction for $\sin^2
\!\theta_{\rm eff}^{\rm lept}$ (second column). The third column indicates
the corresponding requirements for the $R$ measurement. From Ref.~\cite{Jegerlehner:2001wq}.}
\end{center}
\end{table}

As advocated in~\cite{Jegerlehner:1999hg}, the dispersion integral
(\ref{eq:delta_alpha_had}) can be calculated in a different and more
precise  way: it is
sufficient to calculate $\Delta
\alpha^{(5)}_{\mathrm{had}}(s)$ not directly at $s=M_Z^2$, but at some
much lower scale $s_0=-M_0^2$ in the Euclidean region, which is chosen
such that the difference $\Delta\alpha^{(5)}_{\mathrm{had}}(M_Z^2)-
\Delta \alpha^{(5)}_{\mathrm{had}}(-M_0^2)$ can be reliably calculated
using perturbative QCD (pQCD). In (\ref{eq:delta_alpha_had}) pQCD is
used to compute the high energy tail, including some perturbative
windows at intermediate energies. An extended use of pQCD is
possible by monitoring the validity of pQCD via the Adler function,
essentially the derivative of $\Delta
\alpha^{(5)}_{\mathrm{had}}(s)$ evaluated in the spacelike region:
$\frac{D(Q^2)}{Q^2}=-\frac{3\pi}{\alpha}
\,\frac{d \Delta\alpha_{\mathrm{had}}}{d q^2} |_{q^2=-Q^2}$.
Using a 
 state-of-the-art pQCD prediction for the Adler function
one finds that $\Delta\alpha^{(5)}_{\mathrm{had}}(-M_Z^2)-
\Delta \alpha^{(5)}_{\mathrm{had}}(-M_0^2)$ can be neatly calculated from
the predicted Adler function~\cite{Eidelman:1998vc} 
for $M_0 \sim 2.5~\mbox{GeV}$ as a
conservative choice. Also the small missing $\Delta\alpha^{(5)}_{\mathrm{had}}(M_Z^2)-
\Delta \alpha^{(5)}_{\mathrm{had}}(-M_Z^2)$ terms can safely be
calculated in pQCD. The crucial point is that pQCD is used in a fully
controlled manner, away from thresholds and resonances. There are three
points to note: 1) this strategy allows a more precise determination
of $\Delta\alpha^{(5)}_{\mathrm{had}}(M_Z^2)$ than the direct method
based on (\ref{eq:delta_alpha_had}); 2) However, it requires a very
precise QCD calculation and relies on a very precise determination of
the QCD parameters $\alpha_s$, $m_c$ and $m_b$ (for the present status
see \cite{Kuhn:2007vp});
3) Most importantly, as shown in Fig.~\ref{fig:errorprof} {\it down},
the method relies mainly on a precise cross section
measurement in the region below 2.5 GeV, which at the same time is most
important for reducing the uncertainty of the prediction of the muon
$g-2$.  

Projects like KLOE-2 are therefore absolutely crucial for a better
determination of the effective fine structure constant and the muon
$g-2$ (for details see~\cite{Jegerlehner:2008rs}).

%% file: g-2_cb2.tex
\begin{figure}[ht]
\vspace*{-7mm}
\centering
\includegraphics[height=5cm]{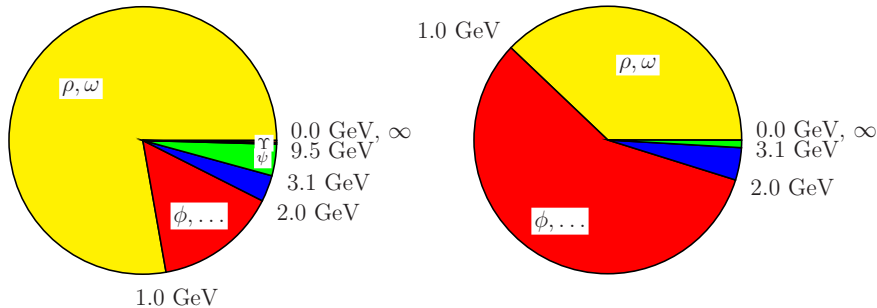}
\caption{The distribution of contributions (left) and errors (right)
in \% for $a_{\mu}^{\mbox{$\scriptscriptstyle{\rm HLO}$}}$ 
from different energy regions. The error of a
contribution $i$ shown is
$\delta^2_{i\:{\rm tot}}/\sum_i \delta^2_{i\:{\rm tot}}$ in \%. The
total error combines statistical and systematic errors in quadrature. From Ref.~\cite{Jegerlehner:2009ry}.}
\label{fig:gmusta}
\end{figure}

Like the effective fine-structure constant at the scale
$M_{\scriptscriptstyle{Z}}$, the {\small SM} determination of the
anomalous magnetic moment of the muon $a_\mu$ is presently limited by the
evaluation of the hadronic vacuum polarisation effects, which cannot be 
computed perturbatively at low energies. 
However, using analyticity and unitarity, it was shown
long ago that the leading-order hadronic contribution to $a_{\mu}$,
$a_{\mu}^{\rm HLO}$, can be computed from hadronic $e^+ e^-$
annihilation data via the dispersion integral~\cite{Bouchiat1961,Gourdin:1969dm}:
\begin{eqnarray}
      a_{\mu}^{\mbox{$\scriptscriptstyle{\rm HLO}$}} &=& 
      \frac{1}{4\pi^3}
      \int^{\infty}_{m_{\pi}^2} {\rm d}s \, K(s) \sigma^{0}\!(s) \nonumber \\
                                                                  &=&
      \frac{\alpha^2}{3\pi^2}
      \int^{\infty}_{m_{\pi}^2} {\rm d}s \, K(s) R(s)/s \, .
\label{eq:amu_had}
\end{eqnarray}
The kernel function $K(s)$ decreases monotonically with increasing~$s$. This
integral is similar to the one entering the evaluation of the hadronic
contribution $\Delta \alpha_{\rm had}^{(5)}(M_{\scriptscriptstyle{Z}}^2)$ in
Eq.~(\ref{eq:delta_alpha_had}). Here, however, the kernel function in the
integrand gives a stronger weight to low-energy data.
The contributions to $a_\mu^{\rm HLO}$ and to its uncertainty $\delta
a_\mu^{\rm HLO}$ from different energy regions
are shown in Fig.~\ref{fig:gmusta}~\cite{Jegerlehner:2009ry}.
The region below 2.0 GeV accounts for about 95\% of the squared uncertainty
$\delta^2 a^{\rm{HLO}}_{\mu}$, 55\% of which comes 
 from the region 1 -- 2 GeV.

In the last few years several papers have been
published~\cite{Hagiwara:2006jt,Jegerlehner:2009ry,Jegerlehner:2008zz,Davier:2009zi,Eidelman:2009ft,Prades:2009qp,Teubner:2010ah} aiming to determine the SM value $a_{\mu}^{\rm
SM}$, including the
evaluation of the $a_\mu^{\rm HLO}$ term based on the 
new measurements of the $e^+e^-$ hadronic cross-sections at low
energy (particularly at VEPP-2M, \Dafne, BEPC, PEP-II and KEKB)\cite{Actis:2010gg}. 
The resulting estimates are systematically lower than 
the experimental result
$     a_{\mu}^{\mbox{$\scriptscriptstyle{\rm exp}$}}= 11659 2080 (63)  
\times 10^{-11}$ ~\cite{Bennett:2006fi} by an amount between 3.1 and 4.0
standard deviations, as for example~\cite{Prades:2009qp}:
\begin{equation}
\Delta a_{\mu} = a_{\mu}^{\mbox{$\scriptscriptstyle{\rm exp}$}}-
a_{\mu}^{\mbox{$\scriptscriptstyle{\rm SM}$}} = +255 (80) \times 10^{-11}.
\end{equation}
\noindent 
As widely discussed in the literature, this result could well 
be the first indirect signal of physics beyond the SM~\cite{Czarnecki:2001pv}.
Deviations of this size are indeed expected in several
realistic ``new-physics'' scenarios, such as the 
minimal supersymmetric extention of the SM
(for a discussion see e.g.~\cite{Stockinger:2006zn,Buchmueller:2009fn,Buchmueller:2007zk}).

 The main contributions to the error on
$a_{\mu}^{\rm SM}$ are shown in Table~\ref{tab:g-2a} for three recent 
estimates~\cite{Jegerlehner:2009ry,Davier:2009zi,Teubner:2010ah}\footnote{Ref.~\cite{Davier:2009zi} includes the recent BaBar $2\pi$ data; \cite{Jegerlehner:2009ry} uses a more conservative error analysis.}, where the two 
dominant contributions to the uncertainty, namely 
$a_\mu^{\rm HLO}$ and the so called hadronic Light-by-Light
scattering term $a_\mu^{\rm LbL}$~\cite{Jegerlehner:2009ry,Prades:2009tw} are shown separately.

\begin{table}[h]
\begin{center}
 \renewcommand{\arraystretch}{1.4}
 \setlength{\tabcolsep}{1.6mm}
{\footnotesize
\begin{tabular}{|c|c c c|c|}
\hline
 Error & \cite{Jegerlehner:2009ry} & \cite{Davier:2009zi} & \cite{Teubner:2010ah} & prospect \\
\hline
 $\delta a_{\mu}^{\rm
SM}$& 65 & 49 & 48 & 35 \\
\hline
 $\delta a_\mu^{\rm HLO}$ & 53 & 41 & 40 & 26 \\
$\delta a_\mu^{\rm LbL}$ & 39 & 26 &26 &  25 \\
\hline
$\delta (a_\mu^{\rm SM} - a_\mu^{\rm EXP})$ & 88 & 80 & 79 & 40 \\
\hline   
\end{tabular}
}
\caption{\label{tab:g-2a} Estimated uncertainties $\delta a_{\mu}$ in units
of $10^{-11}$ according to Refs.~\cite{Jegerlehner:2009ry,Davier:2009zi,Teubner:2010ah} and
(last column) prospects in case of improved precision in the $e^+e^-$
hadronic cross-section measurement (the prospect on $\delta a_\mu^{\rm LbL}$ is an {\it educated guess}).
 Last row: Uncertainty on $\Delta a_{\mu}$ assuming the present experimental error of 63 from BNL-E821~\cite{Bennett:2006fi} (first two
columns) and of 16 (last column) as planned by the future 
(g-2) experiments~\cite{Carey:2009zz,Imazato:2004fy}.}

\end{center}
\end{table}

In order to clarify the nature of the observed discrepancy between theory
and experiment, and eventually
reinforce its statistical significance, 
new direct
measurements of the muon $g-2$ with a fourfold improvement in accuracy 
have been proposed at Fermilab~\cite{Carey:2009zz} and J-PARC~\cite{Imazato:2004fy}.
With these experiments the uncertainty of the difference $\Delta a_{\mu}$ between the experimental and the
theoretical value of 
$a_{\mu}$ will be dominated by the uncertainty of the hadronic
cross sections at low energies, unless new experimental efforts
 at low energy are undertaken.
 The last column of Table~\ref{tab:g-2a} shows
a future scenario based on
realistic improvements in the $e^+e^-\to hadrons$  cross sections measurements. Such
improvements could be obtained by reducing the uncertainties of the hadronic
cross-sections from 0.7\% to 0.4\% in the region below 1 GeV and from 6\%
to 2\% in the region between 1 and 2 GeV as shown in Table~\ref{tab:g-2b}.

\begin{figure}[ht]
\begin{center}
\includegraphics[height=7cm,width=7cm]{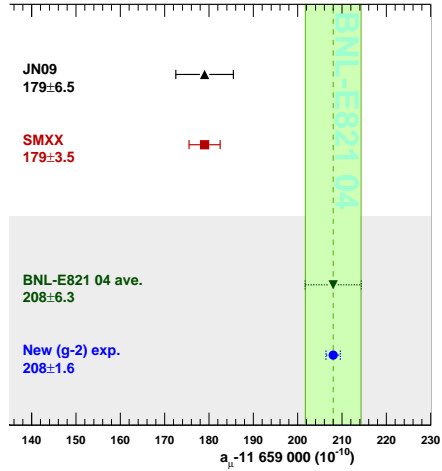}
\caption{Comparison between $a_{\mu}^{\rm SM}$ and $a_{\mu}^{\rm EXP}$. 
``JN09'' is the current evalution of $a_{\mu}^{\rm SM}$ using Ref.~\cite{Jegerlehner:2009ry}; ``SMXX'' is the same central value with a reduced error as
 expected by the improvement on the hadronic cross section measurement at \Dafne-2 (see text); ``BNL-E821 04 ave.'' is the current experimental value
of $a_{\mu}$;  ``New (g-2) exp.'' is the same central value with a fourfold improved accuracy as planned by the future (g-2) experiments~\cite{Carey:2009zz,Imazato:2004fy}.}
\label{fig:g-2xx}
\end{center}
\end{figure}

\begin{table}[h]
\begin{center}
 \renewcommand{\arraystretch}{1.4}
 \setlength{\tabcolsep}{1.6mm}
{\footnotesize
\begin{tabular}{|c|c|c|c|c|}
\hline
 & $\delta (\sigma)/\sigma$ present &$\delta a_{\mu}$present
 & $\delta (\sigma)/\sigma$ prospect &$\delta a_{\mu}$prospect  \\
\hline
$\sqrt{s}<1$~GeV & 0.7\% & 33 & 0.4\% & 19 \\
$1<\sqrt{s}<2$~GeV & 6\% & 39 & 2\% & 13 \\
$\sqrt{s}>2$~GeV & & 12 & & 12 \\
\hline
total & & 53 & & 26 \\
\hline
   
\end{tabular}
}
\caption{\label{tab:g-2b} Overall uncertainty of the cross-section
measurement required to get the reduction of uncertainty on $a_{\mu}$ in units
$10^{-11}$ for
three regions of $\sqrt{s}$ (from Ref.~\cite{Jegerlehner:2008zz}).}
\end{center}
\end{table}
In this scenario the overall uncertainty on $\Delta a_{\mu}$ could be
reduced by a factor 2. In case the central value would remain the same, the
statistical significance would become 7-8 standard deviations, as it can be seen in Fig.~\ref{fig:g-2xx}.

 The effort needed to reduce the uncertainties of the $e^+e^-\to hadrons$
cross-sections according to Table~\ref{tab:g-2b} is challenging but possible, and certainly
well motivated by the excellent opportunity the muon $g$$-$$2$ is
providing us to unveil (or constrain) ``new-physics'' effects.  Once
again, a long-term program of hadronic cross section measurements at low energies is clearly warranted.

%% file: hadc.tex
In the last years the improved precision reached in the
 measurement of $e^+e^-$ annihilation cross sections
in the energy range below a few GeV has led to a substantial reduction 
in the hadronic uncertainty on $\Delta^{(5)}_{\rm{had}}(m_Z^2)$  and 
$a_{\mu}^{\rm{HLO}}$ (as discussed above).
However, while below 1 GeV the error on the two-pion channel
which dominates the cross section in this energy range 
is below 1\%, the region
between 1 and 2 GeV is still poorly known, with a fractional
accuracy of $\sim 6\%$ (see Table~\ref{tab:g-2b}). Since this region contributes about 40\% 
to the total error of the hadronic contribution to the 
effective fine-structure constant to the scale $M_Z$,
$\Delta^{(5)}_{\rm{had}}(m_Z^2)$ (and up to $\sim 70\%$ by using the
Adler function as proposed in~\cite{Jegerlehner:2008rs}), see Fig.~\ref{fig:errorprof}, and $\sim 50\%$ to the error on the
 hadronic contribution of the muon anomaly $a^{\rm{HLO}}_{\mu}$ 
(see Fig.~\ref{fig:gmusta}), it is evident how desiderable an improvement on this region is.

KLOE-2 can play a major role in this region, allowing to measure the hadronic cross section at the 1-2\% level.
With a luminosity of $10^{32}$cm$^{-2}$s$^{-1}$, 
DA$\mathrm{\Phi}$NE upgraded in energy 
can perform a scan in the region from 1 to 2.5 GeV,
collecting an integrated luminosity of 20 pb$^{-1}$ (corresponding
to a few days of data taking) per point. Assuming an energy step of 25 MeV, the whole
region would be scanned in one year of data taking.
\begin{figure}[ht]
\begin{center}
\includegraphics[width=7cm,height=7cm]{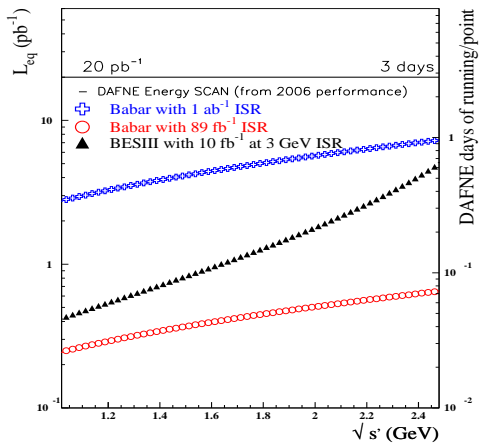}
\vspace{-1.cm}
\caption{\label{fig:w2} Equivalent luminosity for: BaBar with 
1 ab$^{-1}$ 
(cross); BaBar with 89 fb$^{-1}$ (circle); 
BES-III with 10 fb$^{-1}$, using ISR 
 at 3 GeV (triangle). A bin width of 25 MeV is assumed. 
A polar angle of the photon larger than $20^\circ$ is assumed.}
\end{center}
\end{figure}

\begin{figure}[h]
\begin{center}
\includegraphics[width=7cm,height=10cm]{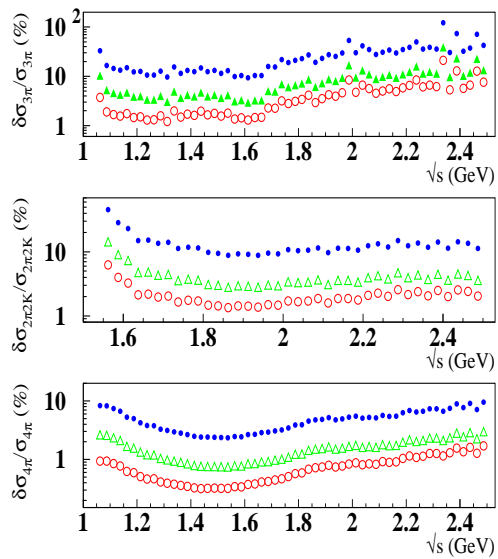}
\vspace{-1.cm}
\caption{\label{fig:impactscan} Comparison of the statistical accuracy in
the cross-section between DA$\mathrm{\Phi}$NE upgraded in energy
 with an energy scan with 20 pb$^{-1}$ per 
point ($\circ$); published BaBar results ($\bullet$), 
BaBar with 890 pb$^{-1}$ statistics (triangle)  for $\pi^+\pi^-\pi^0$
(top), $\pi^+\pi^-K^+K^-$ (middle) and $2\pi^+ 2\pi^-$ (down) channels. 
An energy step of 25 MeV is assumed.}
\end{center}
\end{figure}

As shown in Figure \ref{fig:w2} the statistical yield 
 will be one order of magnitude higher 
than with 1 ab$^{-1}$ at BaBar, and significantly better than BES-III.
%
%
Fig.~\ref{fig:impactscan}  shows the statistical error for the channels
$\pi^+\pi^-\pi^0$, $2\pi^+ 2\pi^-$ and $\pi^+\pi^-K^+K^-$, which can be
 achieved by  an energy scan at DA$\mathrm{\Phi}$NE upgraded in energy
 with 20 pb$^{-1}$ per point,
compared  
with BaBar with published (89 fb$^{-1}$), and tenfold 
(890 fb$^{-1}$) statistics.

As can be seen, an energy scan allows to reach a statistical
 accuracy of the order of  1\% for most of the energy points.
(In addition, KLOE-2 can benefit from the high machine luminosity to
use ISR as well). 
Comparison of exclusive vs inclusive measurements  can be performed as well.

%% file: phys.tex
\subsection{$\gamma\gamma$ physics}
\label{gg}
\input{gg_db2}

\subsection{Spectroscopy and Baryon Form Factors}
\label{spec}
\input{spec_cb}

\subsection{Test of CVC}
\label{cvc}
\input{cvc}

\subsection{Searches}
\label{search}
\input{search_fb}

%% file: gg_db2.tex
\begin{figure}[ht]
\centering
\includegraphics[height=5cm]{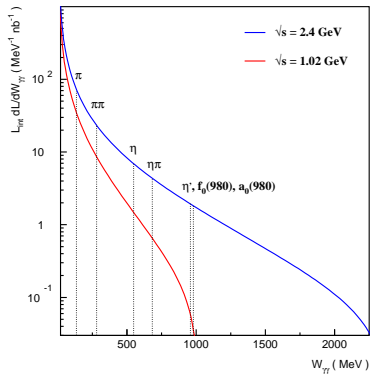}
\caption{Effective $\gamma\gamma$ luminosity as a function of
 $W_{\gamma\gamma}$ corresponding to an integrated luminosity 
 of  1 fb$^{-1}$ at $\sqrt{s} = m_{\phi}$ (red curve) and at 
$\sqrt{s}$=2.4 GeV (blue curve)}
\label{fig:lumgg}
\end{figure}

The upgrade of the KLOE detector with the installation of four smalle-angle 
detectors (taggers)~\cite{Babusci:2009sg} 
for electrons and positrons in the final state of the reaction 
\begin{equation}
e^+e^-\to e^+e^-\gamma^*\gamma^*\to e^+e^- X\;,
\end{equation}
gives the opportunity to investigate $\gamma\gamma$ physics at DA$\Phi$NE. This program will 
benefit from the energy upgrade of DA$\Phi$NE not only for the larger $\gamma\gamma$ flux 
(see Fig.~\ref{fig:lumgg}), but also from the opening of channels not avalaible at the $\phi$ peak, like the production of pseudo-scalar
(like the $\eta'$), scalar ($f_0, a_0, f'_0, a'_0$, etc...),
axial-vector ($a_1, f_1, a'_1, f'_1$, etc...), and tensor ($a_2, f_2$, etc...)
 mesons.

The study of the process in the case in which $X = \pi \pi$ 
is a clean probe to 
investigate the nature of the scalar resonance. The nature of the isoscalar scalars seen in $\pi\pi$ 
scattering below 1.6 GeV, namely the $f_0(600)$ or $\sigma$, $f_0(980)$, $f_0(1370)$ and $f_0(1510)$ 
mesons, is still controversial. Various models have been proposed in which some are $\bar{q}q$, some 
$\bar{qq}qq$, sometimes one is a $\bar{K}K$-molecule, and one a glueball \cite{Klempt:2007cp}, but 
definitive statements cannot be drawn. Their two photon couplings will help unraveling the enigma.

Single pseudoscalar ($X=\pi^0$, $\eta$ or $\eta^\prime$) production is also accessible and would 
improve the determination of the two--photon decay widths of these mesons, relevant for the measurement 
of the pseudoscalar mixing angle $\varphi_P$, and the measurement of the valence gluon content in the
$\eta^\prime$ wavefunction. Moreover, the study of the same processes gives access to the transition form 
factors $F_{X\gamma^*\gamma^*}(M^2_X,q_1^2,q_2^2)$ at spacelike momentum transfers, that are relevant 
for the hadronic Light-by-Light scattering contribution to the $g-2$ of the muon \cite{Jegerlehner:2009ry,Prades:2009qp}~\footnote{Pseudoscalar form factors can be also studied in $e^+e^-\to\gamma^*\to P\gamma$ reactions.}.

\begin{figure}[h]
\begin{tabular}{ccc}
\resizebox{.3\textwidth}{!}{\includegraphics{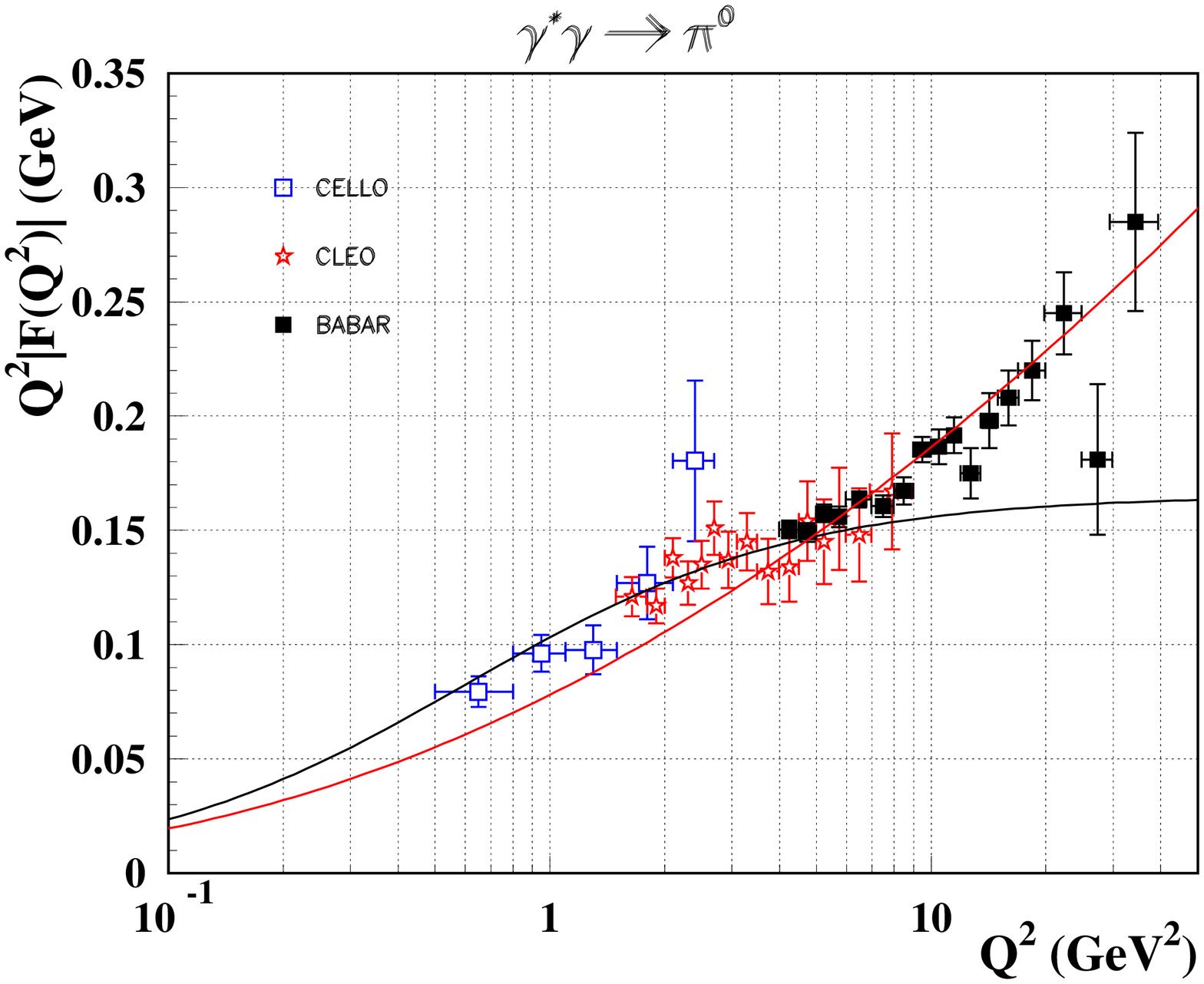}} & 
\resizebox{.3\textwidth}{!}{\includegraphics{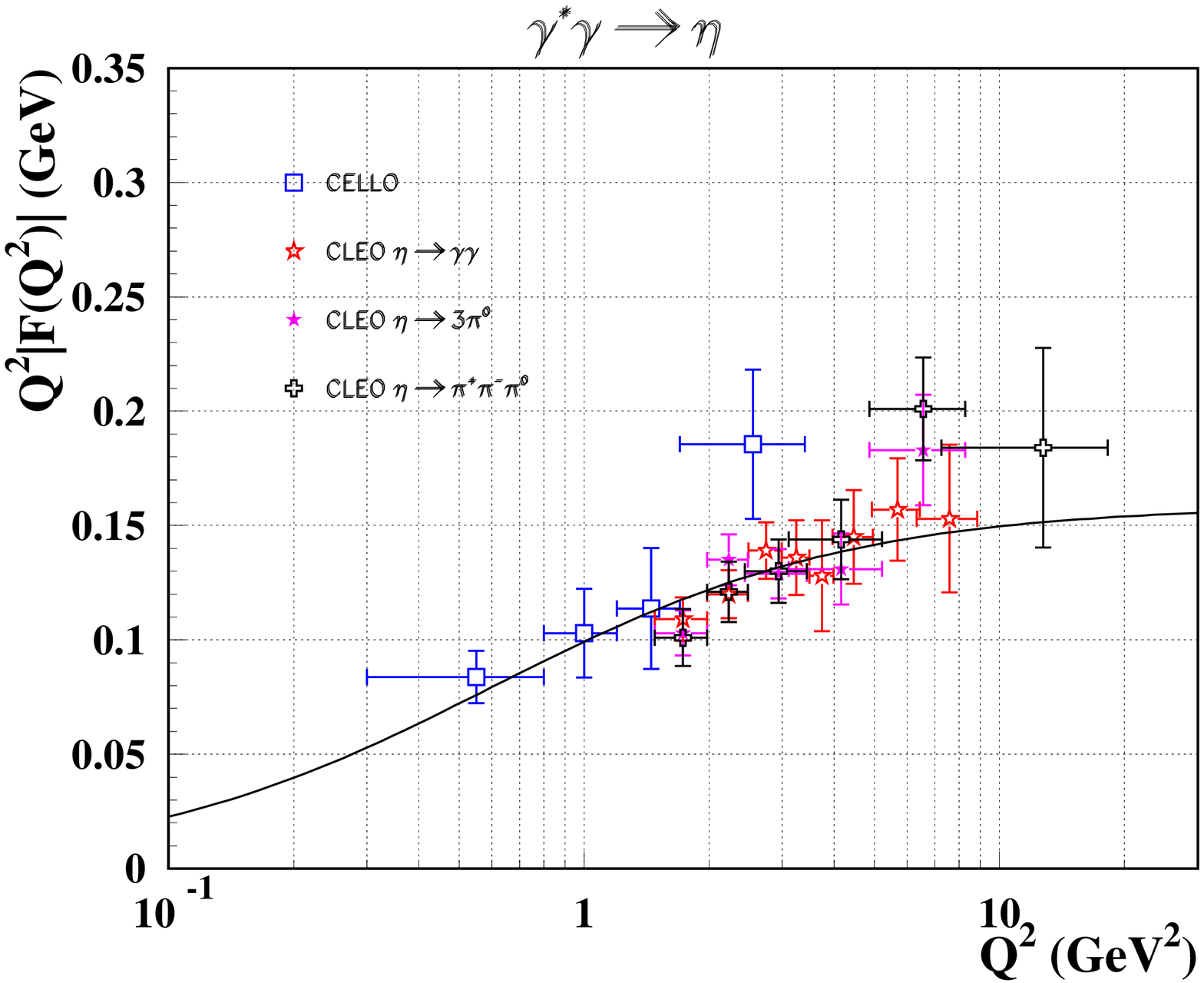}} & 
\resizebox{.3\textwidth}{!}{\includegraphics{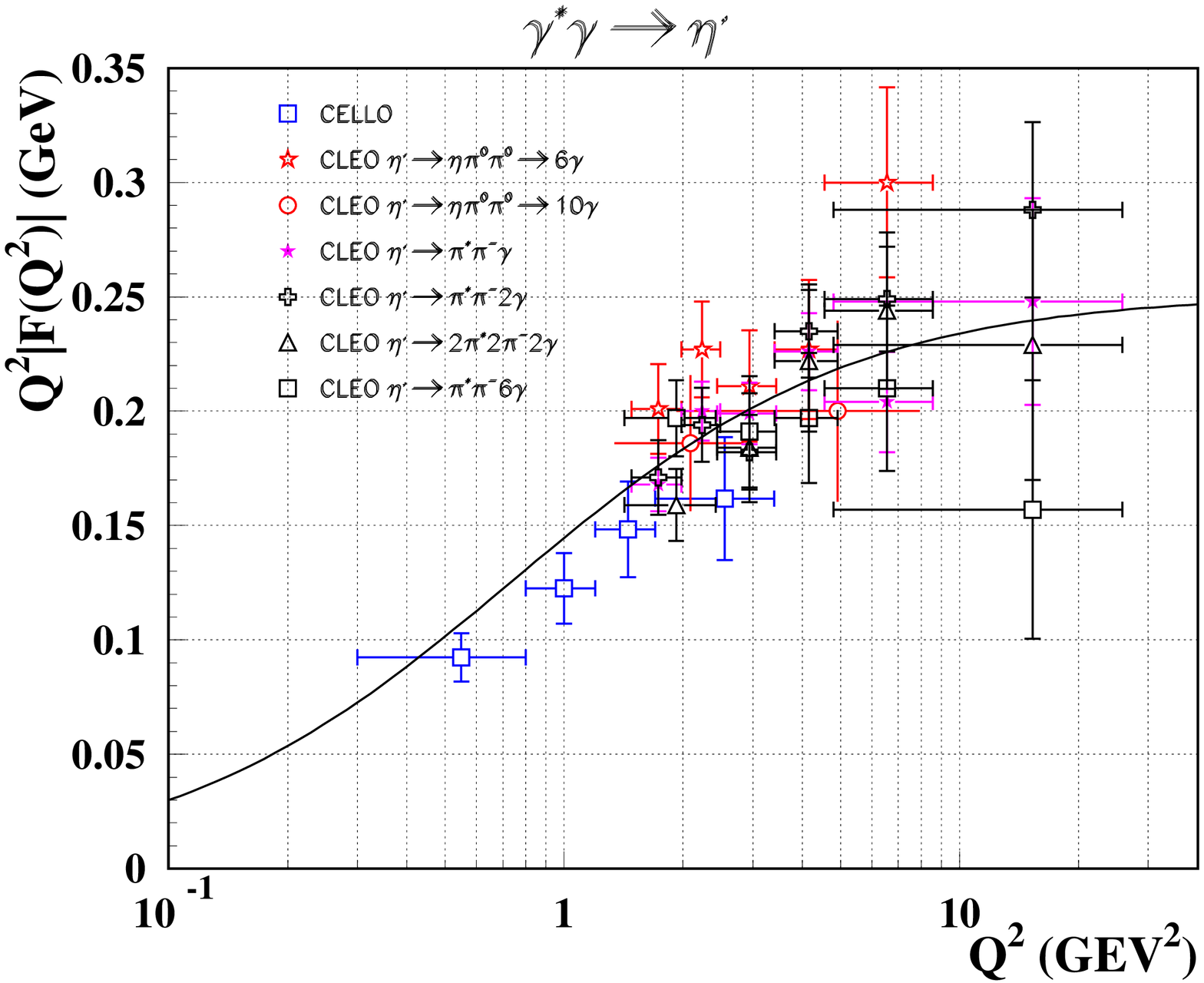}} \\
\end{tabular}
\caption{Left: the $\pi^0$ transition form factor as measured
by the CELLO, CLEO and BaBar experiments. The curve showing an asymptotic 
limit at 160 MeV is from CLEO parametrization \cite{Gronberg:1997fj} 
while the other is from the $F_{\pi\gamma\gamma^*}(m^2_P,Q^2,0)$ 
expression given in Ref. \cite{Dorokhov:2009jd}. Center: the $\eta$ transition 
form factor as measured by CELLO and by CLEO in the specified 
$\eta$ decay channels.
Right: the $\eta^\prime$ form factor as measured by CELLO and by CLEO in the 
specified $\eta^\prime$ decay channels. The curves in the central and 
left panel  
represent the CLEO parametrization 
for the form factors \cite{Gronberg:1997fj}.}
\label{fig:pseudo1}
\end{figure}

By detecting one electron at large angle with respect to the beams, 
the transition form factor $F_{X\gamma\gamma^*}(M^2_X,Q^2,0)$  with one quasi--real and one virtual spacelike photon ($Q^2=-q^2$) can 
be measured. 
These form factors, as reviewed in Ref. \cite{Dorokhov:2009jd}, have 
been measured by the CELLO \cite{Behrend:1990sr}, CLEO \cite{Gronberg:1997fj} and recently BaBar \cite{Aubert:2009mc} 
collaborations in the range $1<Q^2<40$ GeV$^2$ using single--tagged samples. These data are summarized in Fig.~\ref{fig:pseudo1}. The region of very low $Q^2$ (less than  0.5 GeV$^2$, the more important for the Light-by-Light scattering contribution), is devoid of experimental data and is only accessible at DA$\Phi$NE.  

By increasing the  
energy of the machine pseudoscalars (like the $\eta^\prime$), scalars 
(like the $f_0$) and axial-vector (like $a_1$) mesons will be accessible.
 A measurement of all these meson transition form factors will be of 
fundamental importance to reduce the uncertainties that currently affect the estimates of the hadronic LbL scattering.

%% file: spec_cb.tex
Cross sections of exclusive final states are also
important for spectroscopy of vector mesons, whose properties provide
fundamental information on interactions of light quarks. 
PDG lists the following vectors between 1 and 2~GeV~\cite{Yao:2006px}: 
$\omega(1420)$, $\rho(1450)$, $\omega(1650)$, $\phi(1680)$, and
$\rho(1700)$.
However, even their basic parameters ($M,~\Gamma,~\Gamma_{ee}$)
are  badly known. In addition many states still needed a confirmed
 identification.
As discussed in~\cite{kloe2paper,AmelinoCamelia:2010me}
 there are still many 
unsolved points;
 some progress  can be achieved in ISR studies at BaBar and Belle, but 
such analyses are statistically limited, and  a real
breakthrough can happen at the dedicated colliders like
DA$\mathrm{\Phi}$NE-2.

Finally, above a center-of-mass energy of $\sqrt{s}=2M_N=1.88$ GeV,
proton-antiproton and neutron anti-neutron pairs are produced and can be
detected. The measurement of the cross-section for nucleon-antinucleon
pairs allows to extract the nucleon time-like form factors. While the
proton time-like form factors have been extensively measured in a wide
$q^2$ region~\cite{Aubert:2005cb}, the neutron time-like form factors are 
poorly known~\cite{Antonelli:1998fv,Golubev:2009zza}. 
More precise information for both time-like form factors would now
have an important impact on our understanding of the nucleon structure 
(see e.g. Refs.~\cite{Baldini:2007qg,Baldini:2008nk,Lomon:2006xb,Puckett:2010ac,TomasiGustafsson:2005kc} and references therein).


%% file: cvc.tex
The hypothesis of the conserved vector current (CVC) and isospin symmetry 
relate to each other $e^+e^-$ annihilation into isovector hadronic states and corresponding hadronic decays of the $\tau$ lepton~\cite{Tsai:1971vv}.
Using experimental data on $e^+e^-\to$ $hadrons$ with I=1 one can compare the
 CVC predictions and $\tau$ lepton data both for decay spectra and branching ratios.  A systematic check of these predictions showed that at the
(5-10)\% level they work rather well~\cite{Eidelman:1990pb}.
However  new high-precision data on the $2\pi$ final state 
 challenged this statement~\cite{Davier:2002dy,Davier:2003pw,Davier:2009ag} and some evidence for a similar 
discrepancy is also observed in $e^+e^-\to \pi^+\pi^-2\pi^0$~\cite{Davier:2002dy,Davier:2003pw}.
A test of CVC with very high accuracy will be feasible at \Dafne-2 and will require detailed measurements of the energy dependence of the relevant exclusive 
processes, like $\pi^+\pi^-$, $4\pi$ (2 final states), $6\pi$ (3 final states),
$\eta\pi^+\pi^-$, $K_S K_L$ and $K^+K^-$, from threshold to tau lepton mass.


%% file: search_fb.tex
Low energy, high luminosity electron-positron colliders are an ideal 
tool to search for hypothetical ``$U$'' vector bosons weakly coupled with 
Standard Model particles. These bosons are predicted in extensions
of the SM, which have recently appeared in the literature as a consequence
of some intriguing and, as yet not completely explained, astrophysical
observations~\cite{Boehm:2003hm,Pospelov:2007mp,ArkaniHamed:2008qn,Alves:2009nf,Essig:2009nc}.
 In fact KLOE, BaBar and BES-III have already performed or
are planning to perform measurements in the field~\cite{Aubert:2009cp,Aubert:2009cka,Yin:2009mc,Reece:2009un,Bossi:2009uw}.
 There are several 
possible signatures to look at, such as  $e^+e^-\to e^+e^-+\gamma$,
 $e^+e^-\to \mu^+\mu^-+\gamma$, $e^+e^-\to E_{missing}+\gamma$, 
 $e^+e^-\to E_{missing} + e^+e^-$, or events with 4 or 6 leptons in 
the final state. A careful analysis of such reactions in the region 
of interest for this proposal would complement the above mentioned 
searches, particularly  in the case of the channels with missing
energy  or multilepton jets. The cross sections for these processes
are expected to be in the ballpark of 10-100 fb, thus one could 
expect to observe a few hundreds events at the proposed facility.

%% file: dafne.tex
 The possibility to run the \Dafne\
 collider at higher energies has been extensively studied in the 
past~\cite{Milardi:2004mg,danae,pantaleo}. 
In the following the necessary hardware modifications, mode of 
operation and performances estimate will be presented.

\subsection{Injection}

   The injection energy in the \Dafne\ collider is limited by several factors:
\begin{itemize}
\item[-] Linac: the present maximum energy for positrons is about 530 MeV;
\item[-] Damping Ring: the present maximum energy is about 540 MeV;
\item[-] Transfer Lines: the present maximum energy is about 540 MeV;
\item[-] Injection Septa: the present maximum energy is about 540 MeV.
\end{itemize}
   A significative increase of the \Dafne\ injection energy requires 
major changes in the injection complex. In addition a solution to 
inject beams with energy around 1.0 GeV seems unfeasible both for
 the complexity of the changes in the several subsystems and for the
 space constraints that severely limit what can be possibly be adopted.
The most reasonable solution at the moment is to inject in \Dafne\ at 
the nominal energy of about 510 MeV and then ramp the energy up to the 
desired one.

\subsection{Main Rings}

   In order to ramp-up the energy in the main rings the Final Doublet (FD)
 has to be replaced with a Superconducting one. In addition the permanent
 magnet dipole that has been added in the IR this year should also be 
replaced with a superconducting magnet. This solution seems technically 
feasible although a detailed engineering study is necessary before a final 
feasibility statement.
In alternative the FD could be replaced by electromagnetic Panovsky-like
 quadrupoles with a superconducting counter-solenoid to zero the detector 
field on the FD. Also this solution needs detailed studies.

  The maximum beam energy in \Dafne\ is presently determined by the 
Main Dipoles maximum magnetic field and is about 700 MeV. 
A preliminary study of a replacement of these Dipoles has been already made. 
A solution that will allow to reach a beam energy of about 1.02 GeV 
(2.04 CM energy) seems feasible. Again a detailed engineering study 
is necessary before stating the feasibility and the maximum energy 
that can be obtained. This ultimate value could change by +/-10\%
 after such study.

\subsection{Operations and performances}

   The operational scheme would be the following:
\begin{itemize}
\item[-]	Electrons Injection at 510 MeV, followed by the Positron one 
(about 10 minutes total);
\item[-] Ramp up the machine to the desired energy (5 minutes or less);
\item[-] Setting the collisions (2 minutes or less);
\item[-] Coasting in collisions for about 30 minutes;
\item[-] Dump the beams and ramp down the machine (3 minutes).
\end{itemize}

 During the injection and ramping the beams will not be colliding
 (by changing the relative RF phase by 180 degrees) 
and will be enlarged (by increasing the beam coupling by means of
 Skew Quadrupoles) in order to increase the beam lifetimes while 
not making any luminosity. This kind of set-up has been already
 tested in several occasions in MD studies.
    The Wigglers will be OFF all the time since not necessary 
and possibly harmful at higher energy. We foresee a test of 
the maximum storable current without wigglers at 510 MeV 
sometimes in the next run.
    The maximum current will also be somewhat limited by 
the increased Synchrotron radiation power in the Dipoles at 
high energy. At the moment 1-1.5 Amps per beam seems a safe estimate.

    Assuming a peak luminosity of about $5\times 10^{32} cm^{-2} sec^{-1}$ at
 1020 MeV and about 20 pb$^{-1}/day$ average, we should expect about a factor 
2 decrease in the peak luminosity at higher energy 
(because of lower beam-peak currents and possibly not optimal beam parameters)
 and another factor 2 decrease in the integrated luminosity because of
 the duty cycle introduced by the need of ramping. 
An average integrated luminosity of about 5 pb$^{-1}/day$ 
is henceforth estimated which would be allow to scan the region up to $~2$ GeV in about one year (as discussed in Section~\ref{hadc}).

\subsection{Cost estimate and time schedule}

    A realistic cost estimate could be done during the engineering phase
 and probably will not exceed 10 MEur.
 The engineering should be performed with:
\begin{itemize}
\item about one year FTE of engineers expert in magnet design 
\item about three half years FTE of engineers expert in mechanical design, vacuum and cryogenics.
\item about one year FTE of technical support
  \end{itemize}
We estimate that from the date of the approval of this project 
about one year will be needed for the technical design.
 After that about one year will be needed to purchase 
the hardware and about 6 months for the installation in \Dafne.